\documentclass[twocolumn,twoside,slac_two]{revtex4}
\usepackage{aas_macros}

\usepackage{graphicx}
\usepackage{fancyhdr}
\newcommand{\hMsol}{{\>h^{-1}\rm M}_\odot}

\begin{document}

\title{Mapping extragalactic dark matter structures through gamma-rays}

\author{J.~Zavala, V.~Springel and M.~Boylan-Kolchin}
\affiliation{Max-Planck-Institut f\"{u}r Astrophysik, Karl-Schwarzschild-Stra\ss{}e 1, 85740 Garching
bei M\"{u}nchen, Germany}

\begin{abstract}
If dark matter is composed of neutralinos, the gamma-ray radiation produced in
their annihilation offers an attractive possibility for dark matter detection.
This process may contribute significantly to the extragalactic gamma-ray
background (EGB) radiation, which is being measured by the FERMI satellite with
unprecedented sensitivity. Using the high-resolution
Millennium-II simulation of cosmic structure formation we have produced the
first full-sky maps of the expected contribution of dark matter annihilation
to the EGB radiation. Our maps include a proper normalization of
the signal according to a specific supersymmetric model based on
minimal supergravity.  The new simulated maps allow a study of the
angular power spectrum of the gamma-ray background from dark matter
annihilation, which has distinctive features associated with the
nature of the annihilation process. Our
results are in broad agreement with analytic models for the gamma-ray
background, but they also include higher-order correlations not
readily accessible in analytic calculations and, in addition, provide
detailed spectral information for each pixel. In particular, we find
that color maps combining different energies can reveal the cosmic
large-scale structure at low and intermediate redshifts. 
\end{abstract}

\maketitle

\thispagestyle{fancy}

\section{Introduction}

Although dark matter accounts for most of the matter in the Universe, its
nature remains unknown and its presence has only been
inferred through its gravitational effects. However, if dark matter is made 
of neutralinos, a new particle predicted by Supersymmetry, it would also 
interact (although very weakly) with ordinary matter, and it might be detected 
soon in laboratories on Earth. 

In addition, neutralinos can self-annihilate to produce ordinary particles like positrons \cite{Baltz-Edsjo-99}, 
neutrinos \cite{Berezinsky-96} and gamma-ray photons.
If these byproducts of the annihilation are 
copious enough, they could be detected soon.
In the present work, we
will focus on the gamma-ray photons as a residual of the annihilation process.

This gamma-ray radiation is expected to be produced more abundantly
in regions with high dark matter density. 
Thus, it seems best to look for it in very dense nearby regions, such as the 
center of our own Galaxy \cite{Bertone-Merritt-05,Jacholkowska-06} and/or the centers of its satellite
galaxies \cite{Wood-08}. Actually, it turns out that the best prospects for the detection of 
gamma-rays from our Galaxy are obtained by looking slightly off-center to
avoid confusion of the signal with other sources of gamma rays residing at the
Galactic center \cite{Stoehr-03,Springel-08}.

Outside our Galactic halo, gamma-rays are also produced in large 
quantities by the annihilation of dark matter in all the many halos and
subhalos within our past light-cone, contributing to the so-called
extragalactic gamma-ray background (EGB) radiation. The EGB has been measured by
different satellites, in particular in the energy range
between $1\,{\rm MeV}$ and $30\,{\rm GeV}$ by COMPTEL and EGRET
\cite{Strong-Moskalenko-Reimer-04}. Our
understanding of the EGB is greatly improving by the Large Area
Telescope aboard the recently launched FERMI satellite \cite{Atwood-09}, which covers an energy range between
$20\,{\rm MeV}$ and $300\,{\rm GeV}$ and features an improved
sensitivity compared with its predecessor EGRET.

Although the EGB also
receives contributions from other sources, such as blazars \cite{Ando-07a}  and cosmic rays 
accelerated at structure formation shocks \cite{Jubelgas-08}, the energy spectrum and angular
power spectrum of the annihilation radiation have distinctive features that
may open up effective ways for disentangling the signal
\cite{Ando-Komatsu-06,Siegal-Gaskins-Pavlidou-09}. Therefore, a detailed
analysis of the EGB is a viable possibility for detecting dark matter. 

Previous studies have analyzed this possibility using analytic approaches.
However, it is not clear how accurately these methods
capture the non-linear structures resolved in the newest generation of
high-resolution cosmological simulations. Also, the previous analysis
have so far been restricted to statistical statements about the power
spectrum of the EGB, or its mean flux. For a full characterization of
the signal, it would be useful to have accurate realizations of maps
of the expected gamma-ray emission over the whole sky. 

In our study \cite{Zavala-Springel-Boylan-Kolchin-09}, we therefore focus on predicting the
EGB radiation directly from cosmological N-body
simulations. To this end, we use the Millennium-II simulation (MS-II) \cite{Boylan-Kolchin-09}. With
$10^{10}$ particles in a homogeneously sampled volume of
$(100\,h^{-1}{\rm Mpc})^3$, it is one of the best resolved structure
formation simulations to date. We use a ray-tracing technique to
accumulate the signal from all halos and subhalos over the past
light-cone of a fiducial observer, positioned at a plausible location of
the Milky Way in the simulation box. 

The values we calculate for each
pixel of such a sky map are that of the specific intensity, the
energy of photons received per unit area, time, solid angle and energy
range:
\begin{equation}\label{intensity}
I_{\gamma,0}=\frac{1}{4\pi}\int
\epsilon_{\gamma,0}(E_{\gamma,0}(1+z),z)\frac{{\rm d}r}{(1+z)^4},
\end{equation}
where the integral is over the whole line of sight, $r$ is the
comoving distance and $E_{\gamma,0}$ is the energy measured by the
observer at $z=0$. The gamma ray emissivity $\epsilon_{\gamma}$ (energy
emitted per unit energy range, unit volume and unit time) associated
with dark matter annihilation is given by:
\begin{equation}\label{emiss}
\epsilon_{\gamma}=E_{\gamma}\frac{{\rm d}N_{\gamma}}{{\rm d}E_{\gamma}}\frac{\langle\sigma
v\rangle}{2}\left[\frac{\rho_{\chi}}{m_{\chi}}\right]^2,
\end{equation}
where  ${\rm d}N_{\gamma}/{\rm d}E_{\gamma}$  is the  total  differential photon  spectrum
summed  over all  channels  of annihilation, $m_{\chi}$ and $\rho_{\chi}$ are
the mass and density of neutralinos, and $\langle\sigma v\rangle$ is the thermally averaged product of the
annihilation cross section and the M\o ller velocity. Note that $\epsilon_{\gamma,0}$ is evaluated at the
blueshifted energy $(1+z)E_{\gamma,0}$ along the line-of-sight to
compensate for the cosmological redshifting. 

\section{Dark matter annihilation and the SUSY factor}

Of  all the  quantities in  Eq. (\ref{emiss}),  only $\rho_{\chi}$  depends  on the  spatial  distribution of
neutralinos, the rest is related to its intrinsic properties, which we can
conveniently analyze separately by defining the so called SUSY factor:
\begin{equation} \label{fsusy}
f_{\rm SUSY}= \frac{{\rm d}N_{\gamma}}{{\rm
    d}E_{\gamma}}\frac{\langle\sigma v\rangle}{m^2_{\chi}} .
\end{equation}
In the present section we describe the particular SUSY model that we
used to compute the gamma-ray emissivity produced by dark matter
annihilation.

We restrict our analysis to the framework of the
minimal supergravity model (mSUGRA), which is popular, among
other reasons, due to its relative simplicity. The
large number of free parameters in general SUSY is reduced
to effectively four free parameters in mSUGRA: $m_{1/2}$, $m_0$, $A_{0}$, which are 
the values of the gaugino and scalar masses and
the trilinear coupling, all specified at the GUT scale, 
tan$\beta$ which is
the ratio of the expectation values in the vacuum of the two neutral
SUSY Higgs, and finally, the sign of $\mu$,
the Higgsino mass parameter.

The general 5-dimensional parameter space of the mSUGRA model is
significantly constrained by various requirements: consistency with
radiative electroweak symmetry breaking (EWSB) and experimental
constraints on the low energy region. It is also normally assumed that
if the major component of dark matter is the lightest neutralino, then its relic density
$\Omega_{\rm \chi}$ should be equal to the observed abundance of dark matter 
$\Omega_{\rm DM}$. This condition reduces the
allowed regions in the parameter space considerably. 

The different regions in the mSUGRA parameter space that satisfy these
constraints have been studied often in the past
and have received generic names: (B) bulk region (low values of $m_0$ and $m_{1/2}$), (FP)
focus point region (large values of $m_0$), (CA) co-annihilation region
(low $m_0$ and $m_{\chi}\lesssim m_{\tilde{\tau_1}}$, where
$\tilde{\tau_1}$ is the lightest slepton) and (RAF) rapid annihilation
funnel region (for large values of tan$\beta$ and a specific condition
for $m_{\chi}$\footnote{The relation which should hold is:
  $2m_{\chi}\approx m_{A}$, where $A$ is the CP-odd Higgs boson; see
  for example \cite{Feng-05} for details.}). Instead of making a full scan of
the allowed parameter space, we define several
``benchmarks points'' which fulfill all the experimental constraints
discussed above and which can be taken as representative of the
different regions that we just described. The
benchmark points were selected following the analysis of
\cite{Battaglia-04} and \cite{Gondolo-04}.
We use the numerical code {\small DarkSUSY}
\cite{Gondolo-04,Gondolo-05} to analyze these benchmark points. 

\begin{figure}
\includegraphics[height=7.5cm,width=7.5cm]{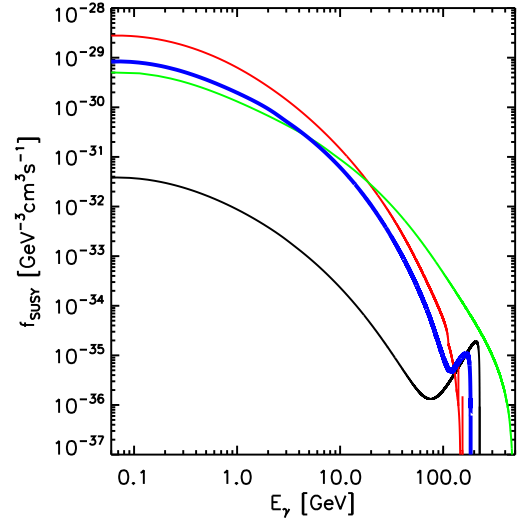}
\caption{{\footnotesize SUSY factor as a function of $\gamma$-ray 
  emission energy for a selection of benchmark points representatives of
  the allowed regions in the mSUGRA parameter space: CA (black), FP (red), RAF (green ) and B (blue).}}
\label{susy_energy}
\end{figure}

Fig.~\ref{susy_energy} shows the energy spectrum of $f_{\rm SUSY}$
computed for four of the benchmark points analyzed, which are representatives
of the different regions: CA (black), FP (red), RAF (green ) and B (blue). The main features of
Fig.~\ref{susy_energy} are determined by the photon spectrum ${\rm
  d}N_{\gamma}/{\rm d}E_{\gamma}$ which receives contributions from
three mechanisms of photon production \cite{Bringmann-Bergstrom-Edsjo-08}: 
(i) gamma-ray continuum emission following the decay of neutral pions produced during the
hadronization of the primary annihilation products;
(ii) monoenergetic gamma-ray lines for neutralino annihilation
in two-body final states containing photons; (iii) internal
bremsstrahlung (IB), which leads to the emission
of an additional photon in the final state; the IB contribution to the
spectrum is typically dominant at high energies resulting in a
characteristic bump. Processes (ii) and
(iii) are subdominant to process (i), but they display
distinctive spectral features intrinsic to the phenomenon of
annihilation. 

In Fig.~\ref{susy_energy}, we have shown only 4 of all the benchmark
points we have considered; the rest have very similar features and lie in
between these 4 cases. Fig.~\ref{susy_energy} also indicates that the
normalization of the spectrum of $f_{\rm SUSY}$ typically varies by
three orders of magnitude among the different regions of the allowed
parameter space.

For the rest of our analysis, we will use one particular
benchmark point, which is
the one we highlight with a thick blue line in
Fig.~\ref{susy_energy}. It is a model with $m_{\chi}=185$~GeV
that gives an upper limit on $f_{\rm SUSY}$
for the benchmark points in the bulk region and is also close to the
maximum value of $f_{\rm SUSY}$ for all the benchmark points analyzed.
The latter is the main reason to choose this particular model because
it is desirable that the predicted gamma-ray flux
has a large value among the different theoretical possibilities. There are other
reasons as well. The mass of the neutralino in this model is
``safely'' larger than the lower mass bounds coming from experimental
constraints ($m_\chi>50\,{\rm GeV}$ according to \cite{Heister-04}),
but low enough to be detectable in the experiments available in the
near future. Also, a high value of the parameter tan$\beta$ seems to
be favored by other theoretical expectations \cite{Nunez-08}.

\section{The dark matter annihilation contribution to the EGB}

The specific intensity (Eq.~\ref{intensity}) describes the total
emission from dark matter annihilations integrated over the full
backwards light cone along a certain direction.  We use the data of
the MS-II to fill the whole volume contained in the past-light cone of
an observer located at a fiducial position at $z=0$. Since there
are only 68 simulation outputs in total, we can approximate the temporal
evolution of structure growth by using at each redshift along the past
light-cone the output time closest to this epoch. To cover all space,
we use periodic replication of the simulation box. 

Our strategy is not to account for the emission
at the level of individual dark matter particles, but instead to use entire halos
and subhalos, allowing us to accurately correct for resolution
effects. The list of dark matter particles at each output time is
hence replaced by a catalog of dark matter substructures. 

We further assume
that these substructures are well represented by a spherically symmetric NFW
density profile \cite{Navarro-Frenk-White-97}. Under this assumption the
gamma-ray luminosity from a given substructure can be written as a scaling law
\cite{Springel-08}:
\begin{equation}\label{scaling_law}
L_h=\int\rho^2_{\rm NFW}(r)\,{\rm d}V=\frac{1.23\,V_{\rm
    max}^4}{G^2r_{\rm max}},
\end{equation}
where $V_{\rm max}$ is the maximum rotational
velocity of the subhalo and $r_{\rm max}$ is the radius where this maximum is reached.
The most recent high-resolution N-body simulations support a
slightly revised density profile that becomes gradually shallower
towards the center, the so called Einasto profile. For this case, the
coefficient $1.23$ in Eq.~(\ref{scaling_law}) changes to 1.87, increasing our
results effectively by 50$\%$. This is an
insignificant difference compared to other uncertainties in the
analysis of the dark matter annihilation radiation. For the purposes of our work, the NFW profile is
a good approximation and using it simplifies comparisons to a number of
results in the literature.

Our map-making method then becomes a task to accurately
accumulate the properly redshifted emission from these structures in
discretized representations of the sky and to include a resolution correction for poorly
resolved objects.  All
our maps use $N_{\rm pix}=12\times 512^2\sim \pi \times 10^6$ pixels,
corresponding to an angular resolution of $\sim 0.115^\circ$.

A given pixel in the simulated maps covers a solid angle
$\Delta\Omega_{\rm pix}$. Our map-making code computes the average
value of the specific intensity within the area subtended by this
solid angle by conservatively distributing the emission of each
substructure over the appropriate pixels.  Combining
Eqs.~(\ref{emiss}$-$\ref{scaling_law}) we can write the relevant sum
over all substructures in the light-cone as:
\begin{eqnarray}\label{intensity_pix}
I_{\gamma,0}(\Delta\Omega_{\rm pix})=\ \ \ \ \ \ \ \ \ \ \ \ \ \ \ \ \ \ \ \ \ \ \ \ \ \ \ \ \ \ \ \ \ \ \ \ \nonumber\\
\frac{1}{8\pi}\sum_{h\epsilon\Delta\Omega_{pix}}L_{h}w(d_h,r_h)E_{\gamma,0}f_{\rm
  SUSY}(z_h)\vert_{E_{\gamma,0}},
\end{eqnarray}
where the function $w(d_h,r_h)$ is a weight function that distributes
the luminosity of a given halo onto the pixels overlapping with the
projected ``size''of the
halo; the latter depends on the distance of the observer to the halo
$d_h$ and the transverse distance $r_h$ between the halo center and
the center of the pixels it touches. Except for structures that are
very nearby, the high central concentration of the emission of a
subhalo and the limited angular resolution of our maps give most
subhalos the character of unresolved point sources.

For the purpose of extending the predictions of our maps down to the
damping scale limit of neutralinos ($\sim10^{-6}$M$_{\odot}$), we 
divide the maps into
separate components: (i) the contribution of {\em resolved halos and
  subhalos} with a minimum mass of $6.89\times10^8\hMsol$ (called
``ReHS'' in the following); (ii) the contribution of {\em unresolved
  main halos} with masses in the range $1.0\times
10^{-6}-6.89\times10^8\hMsol$ (referred to as ``UnH''); and finally,
(iii) the contribution of {\em unresolved subhalos} in the same mass
range as in the case (ii) (component ``UnS'').

\subsection{Resolved halos and subhalos (ReHS)}

\begin{figure*}
\resizebox{11cm}{!}{\includegraphics{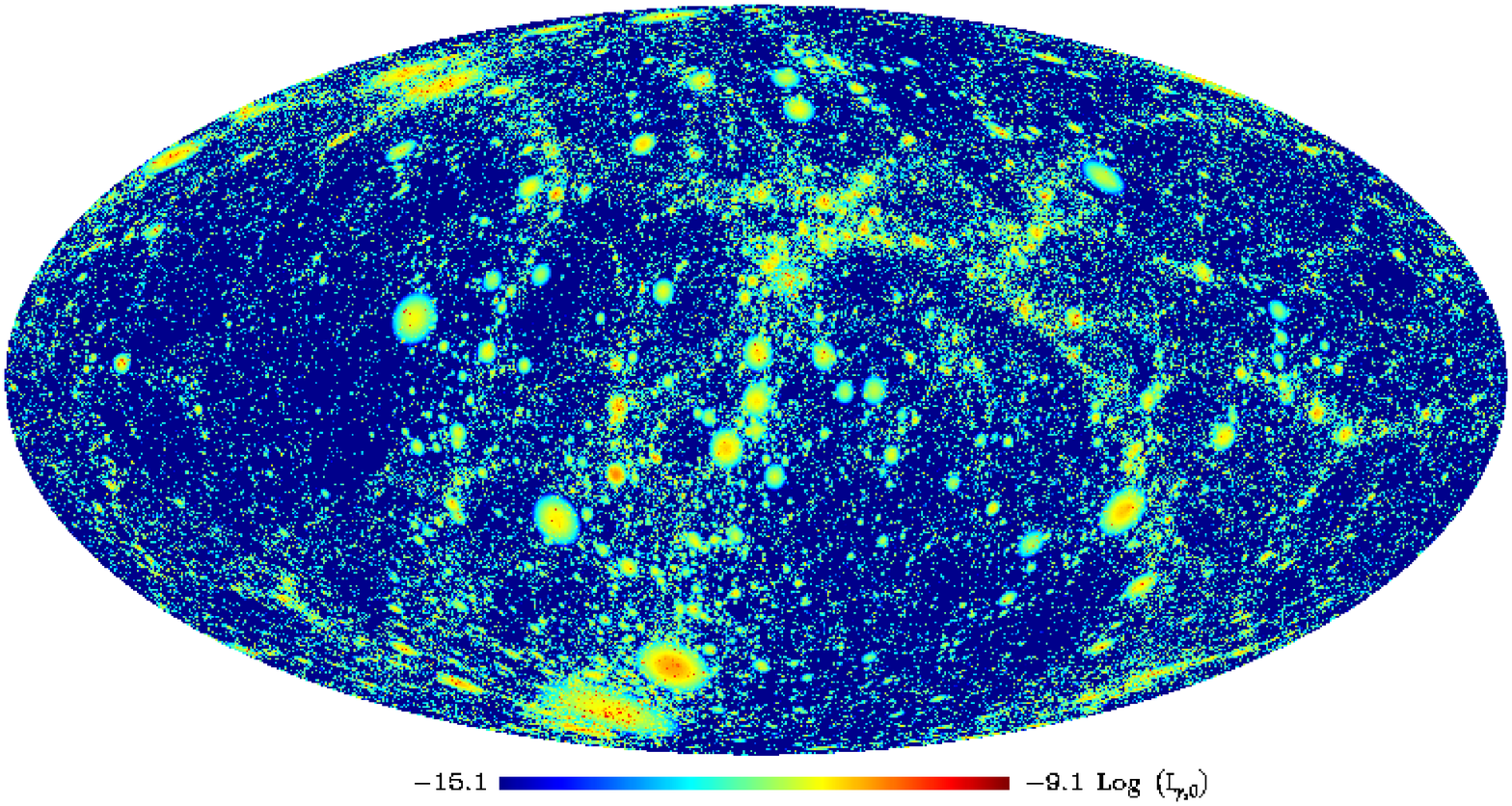}}
\resizebox{11cm}{!}{\includegraphics{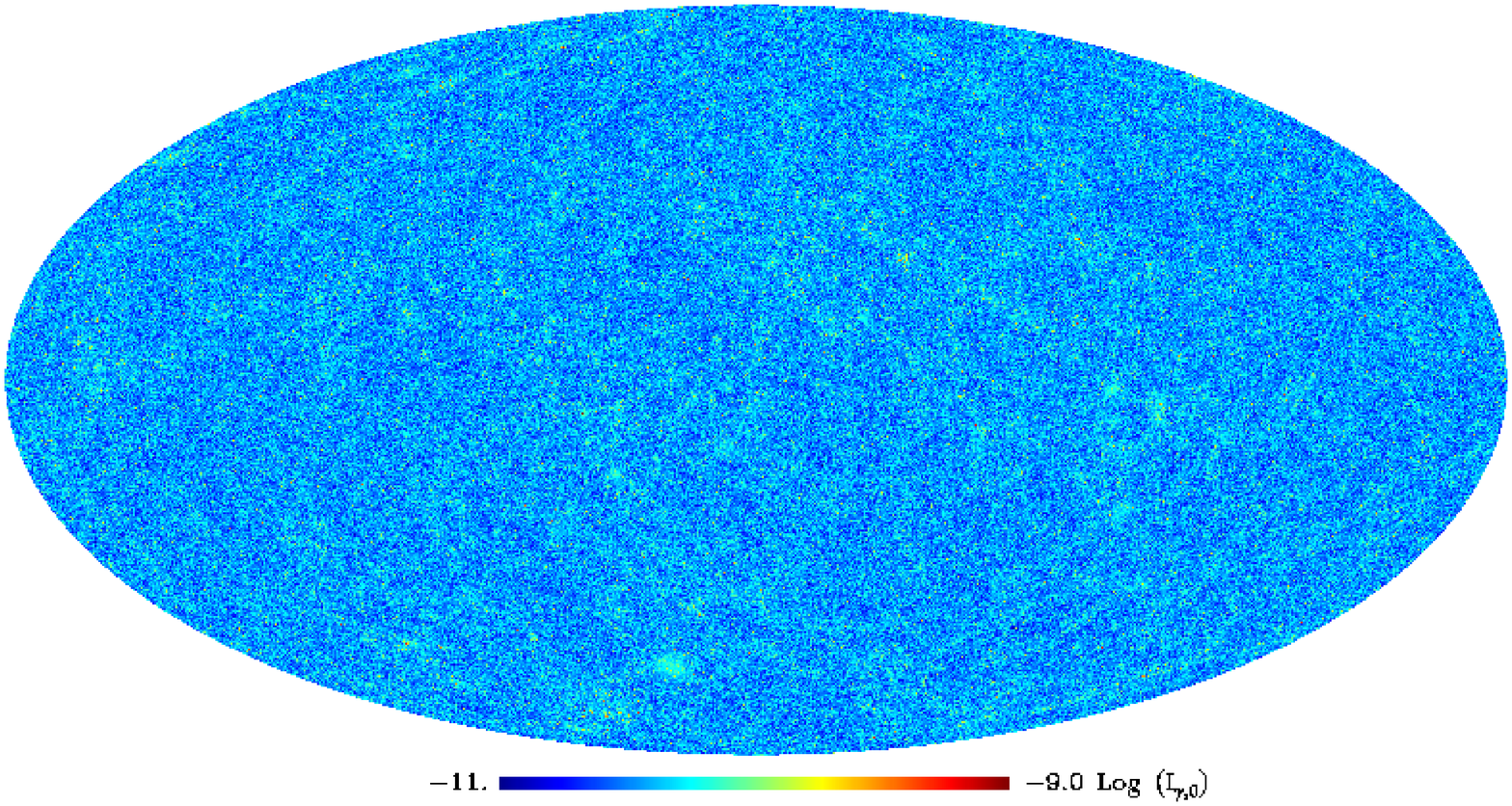}}
\caption{{\footnotesize {\it Upper panel:} A partial map showing
  the EGB produced by dark matter
  annihilation in nearby structures. Only sources within 68 Mpc from the
  observer are considered for the
  map. The color scale gives a visual impression of the
  values of the specific gamma-ray intensity for each pixel in the map; the red
  color corresponds to the highest values of specific intensity.  The
  observed energy of the simulated gamma-ray radiation is 10 GeV.  
  {\it Lower panel:} The full gamma-ray sky map from dark matter
  annihilation containing sources up to $z\sim10$. In both maps only the contribution of the
  smallest halos resolved by the simulation ($\sim10^9\hMsol$)
  has been taken into account.}}
\label{map}
\end{figure*}

In Fig.~\ref{map}, we show full sky maps of the the $\gamma$-ray emission of
all main halos and subhalos that are detected in the MS-II; this corresponds
to an essentially perfectly complete sample of all halos above a mass limit of
$6.89\times10^8\hMsol$. In the top panel of Fig.~\ref{map}, a partial map at
low redshift (corresponding to the first shell at $z=0$) is shown in order to illustrate typical foreground
structure, whereas the bottom panel gives an integrated map out to $z=10$,
which is approximately the full EGB from annihilation.  The maps in
Fig.~\ref{map} are for $E_{\gamma,0}=10\,{\rm GeV}$. In the color scale used for the
maps, red corresponds to the highest and black to the lowest values of the
specific intensity.

\subsection{Unresolved halos (UnH)}

The maps shown in Fig.~\ref{map} are complete down to the minimum mass in the
MS-II that we can trust, $\sim6.89\times10^8\hMsol$. In order to make a
prediction of the full EGB coming from dark matter annihilations, we
extrapolate the $\gamma$-ray flux to account for the contribution of all
missing dark matter halos down to the cutoff mass $10^{-6}\hMsol$. For
this purpose we use extrapolations of the power law behavior found for the
total gamma-ray luminosity of host halos as a function of mass. This is an analysis that we present in
detail in section 4.1 of \cite{Zavala-Springel-Boylan-Kolchin-09}.

The way we incorporate this extrapolated contribution in the $\gamma$-ray maps is the
following.  We assume that the EGB radiation from the missing halos in the
mass range $10^{-6}\hMsol$ to $\sim6.89\times10^8\hMsol$ is distributed on
the sky in the same way as the one from the smallest masses we can resolve in
the simulation, which we adopt as the mass range between
$1.4\times10^{8}\hMsol$ and $\sim6.89\times10^8\hMsol$.  Hence, we compute the value of
a boost factor $b_h$ with which each resolved halo in the mass range
$1.4-6.89\times10^8\hMsol$ needs to be multiplied such that the luminosity of
the unresolved main halos is accounted for as well. We found that
$b_h\sim60$, and that its value is nearly independent of redshift up to the
highest redshift we can reliable make the extrapolation, which is $z\sim2.1$.

\subsection{Unresolved subhalos (UnS)}

To add the contribution of unresolved substructures to the $\gamma$-ray maps,
we first assume that the identified subhalo population is complete down to a
mass limit of $M_{\rm sub}=6.89\times10^8\hMsol$. Below this mass we compute
the gamma-ray luminosity of subhalos using extrapolations of the power law behavior
of the contribution of these subhalos to the luminosities of their host
halos. The parameters of the power law are given in detail in section 4.2 of
\cite{Zavala-Springel-Boylan-Kolchin-09}. We add this contribution in different ways for the
following two cases.

In the case of resolved main halos with more than 100 particles, 
we distributed the extra luminosity coming from unresolved substructures
among the subhalos of the main host, assuming in this way that unresolved
subhalos are distributed in the same way as the resolved ones. If the
main halos have no subhalos, the extra luminosity is given directly to them. 

For the second case of all main halos with masses
less than $6.89\times10^8\hMsol$, we use the following strategy.  From the
results described in the 
last subsection we can get the total luminosity coming from all
main halos with masses between the damping scale limit and
$6.89\times10^8\hMsol$. Thus,
we can compute the boost factor $b_{\rm sub}$ for which this luminosity needs
to be multiplied in order to include the full contribution of all subhalos of the
main halos in this mass range. We found that $b_{\rm sub}$ is roughly independent of redshift. Using the
range of values found for the power laws described before, we obtain that
$b_{\rm sub}\approx2-60$.  We take the extreme values
of this range in the following results as the minimum
and maximum values that reflect the uncertainties in our extrapolation method and should
bracket the true result.

\subsection{Isotropic and anisotropic components of the EGB from dark matter
  annihilation}

\begin{figure}
\includegraphics[height=9cm,width=7.8cm]{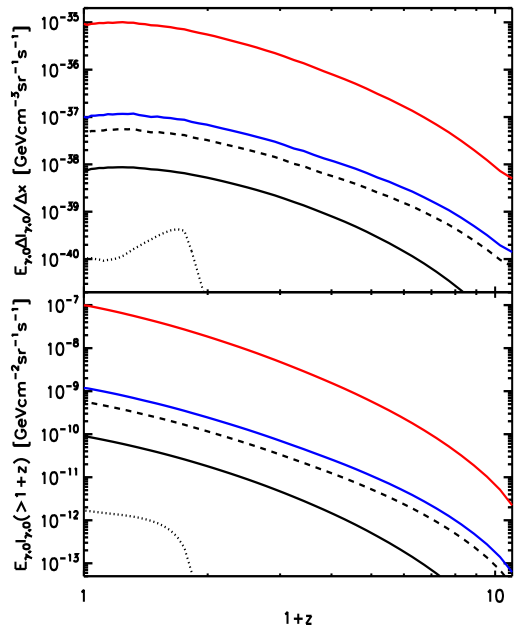}
\caption{{\footnotesize {\it Upper panel: }Mean annihilation intensity per comoving
  shell thickness, as a function of
  redshift for individual partial maps. The solid black line is for
  resolved halos and subhalos. The dashed black line is for main
  halos only, resolved and unresolved down to the cutoff mass
  using a boost factor $b_h=60$. The blue and red solid lines show the
  contribution of all components, resolved and unresolved halos and
  subhalos, boosted with the extreme values in the interval $b_{\rm
    sub}=2-60$.  All lines are for $E_{\gamma,0}=10\,{\rm GeV}$,
  except for the black dotted line which is for the ReHS component with 
  $E_{\gamma,0}=100\,{\rm GeV}$. {\it Lower panel: } The same as the
  upper panel but for the accumulated intensity,
  $I_{\gamma,0}(>1+z)$.}}
\label{maps_means}
\end{figure}

The upper panel of Fig.~\ref{maps_means} shows the mean value of the specific
intensity $\Delta I_{\gamma,0}$ divided by the comoving thickness of the
shell at $E_{\gamma,0}=10\,{\rm GeV}$ as a function of redshift. Recall that each partial map represents the
total specific intensity in a shell of constant comoving thickness, so this
provides information about where the signal is coming from.  The black line is
for the ReHS case, while the dashed line is for main halos only (resolved and
unresolved, including the boost factor $b_h=60$). The remaining lines are for
the resolved and unresolved main halos and their subhalos, boosted to
include the contribution of structures all the way down to the damping scale
limit; here the blue line is for the ``minimum boost'' with $b_{\rm sub}=2$
whereas the red line is for the ``maximum boost'' with $b_{\rm sub}=60$, but
both include the boost $b_h=60$ for unresolved main halos.

In all cases, the contribution from partial maps up until $z\sim1$ is almost
constant, indicating that the increasing number of sources for shells at
higher redshifts, due to the larger volume seen behind each pixel, is
approximately compensated by the distance factor and the spectral effects from
the cosmological redshifting and the intrinsic variation of the emission
spectrum, as described by $f_{\rm SUSY}$ (see Fig.~\ref{susy_energy}).

The lower panel of Fig.~\ref{maps_means} shows the accumulated mean value of
the specific intensity as a function of redshift. This clearly shows that the
most relevant contributions come from maps up to $z\sim2$. Overall,
applying the maximum boost increases the mean value of $I_{\gamma,0}$ of the
ReHS case by three orders of magnitude.  For most energies,
the $f_{\rm SUSY}$ spectrum is monotonically decreasing. In these regions, the
dependence on $I_{\gamma,0}$ with redshift will have the generic form shown in
Fig.~\ref{maps_means}. However, at the highest energies the shape is expected
to change dramatically due to the importance of Internal Bremsstrahlung and/or
of monochromatic lines (see Fig.~\ref{susy_energy}) close to the rest mass
energy of the dark matter particle. Such an effect is clearly visible in the
black dotted line in Fig.~\ref{maps_means}, which is the ReHS case for an
energy of $100\,{\rm GeV}$ and features a bump in the redshift distribution, a
reflection of the importance of IB at the highest energies.

The angular power spectrum of the EGB is an important tool to study the
statistical properties of its anisotropy on the sky. In fact, certain classes
of $\gamma$-ray sources are expected to exhibit different power spectra,
making this a potential means to identify the origin of the unresolved EGB.

\begin{figure}
\includegraphics[height=7.5cm,width=7.5cm]{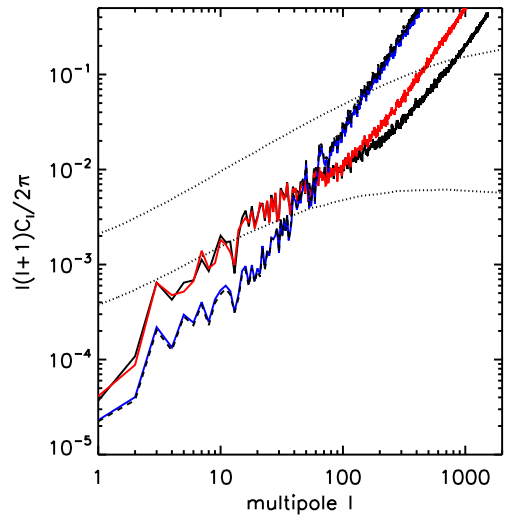}
\caption{{\footnotesize Angular power spectrum of the EGB produced by dark matter
  annihilation as a function of the multipole $l$ for $E_{\gamma,0}=10\,{\rm
    GeV}$. The line styles are as in Fig.~\ref{maps_means} except for the
  dotted lines that show
  the predictions from \citet{Ando-07b} for a subhalo-dominant contribution
  with and without considering tidal destruction (lower and upper
  dotted lines), respectively.}}
\label{power}
\end{figure}

In Fig.~\ref{power}, we show the angular power spectrum $l(l+1)C_l/2\pi$ as
a function of the multipole $l$, at an observed energy of
$E_{\gamma,0}=10\,{\rm GeV}$, for all the cases discussed in
Fig.~\ref{maps_means}. At large scales, $l\lesssim10$, the power spectrum is
related to the clustering of dark matter halos. When only the main halo contribution to the EGB is
considered (dashed black line), the normalization is lower because most of the
$\gamma$-ray signal comes from low mass halos that are less clustered
than more massive halos. 

The blue line in Fig.~\ref{power}, corresponding to the full extrapolation
including subhalos but with the minimal boost $b_{\rm sub}=2$, has exactly the
same power spectrum than the case with main halos only, at all scales.  This
is because the signal from main halos is dominant in this case, and subhalos
have a negligible effect. In contrast, in the case where subhalos have a
significant contribution mediated by $b_{\rm sub}=60$ (red line), the
normalization is larger at small scales because the signal is dominated by
subhalos belonging to the most massive halos, and the latter are strongly
clustered.  

As the angular scale decreases, $l>10$, the power spectrum depends more and
more on the internal structures of halos. For the cases where substructures
are ignored or are negligible (dashed-black and blue lines), the slope becomes
steeper, with a slope close to 2. The power spectrum for the cases where
substructures are relevant (red and black solid lines) behaves differently,
however.  In the range $l\in[20,100]$, it becomes slightly shallower, i.e.~the
signal is slightly more isotropic in this regime.  Contrary to
the strong central concentration of the matter in a halo, the number density
profile of subhalos is considerably shallower than a NFW profile and produces
a luminosity profile in projection which is essentially flat
\cite{Springel-08}.  This effect continues until $l\sim200$ where the power
spectrum becomes dominated by the low-mass main halos.  

The dotted lines in Fig.~\ref{power} were taken from the analytical results of
\cite{Ando-07b} and are shown here for comparison, see the las paragraphs of
section 5.5 in \cite{Zavala-Springel-Boylan-Kolchin-09} for a discussion on the differences between
our results and those of \cite{Ando-07b}.

\begin{figure*}
\centering                      
\includegraphics[width=0.49\hsize]{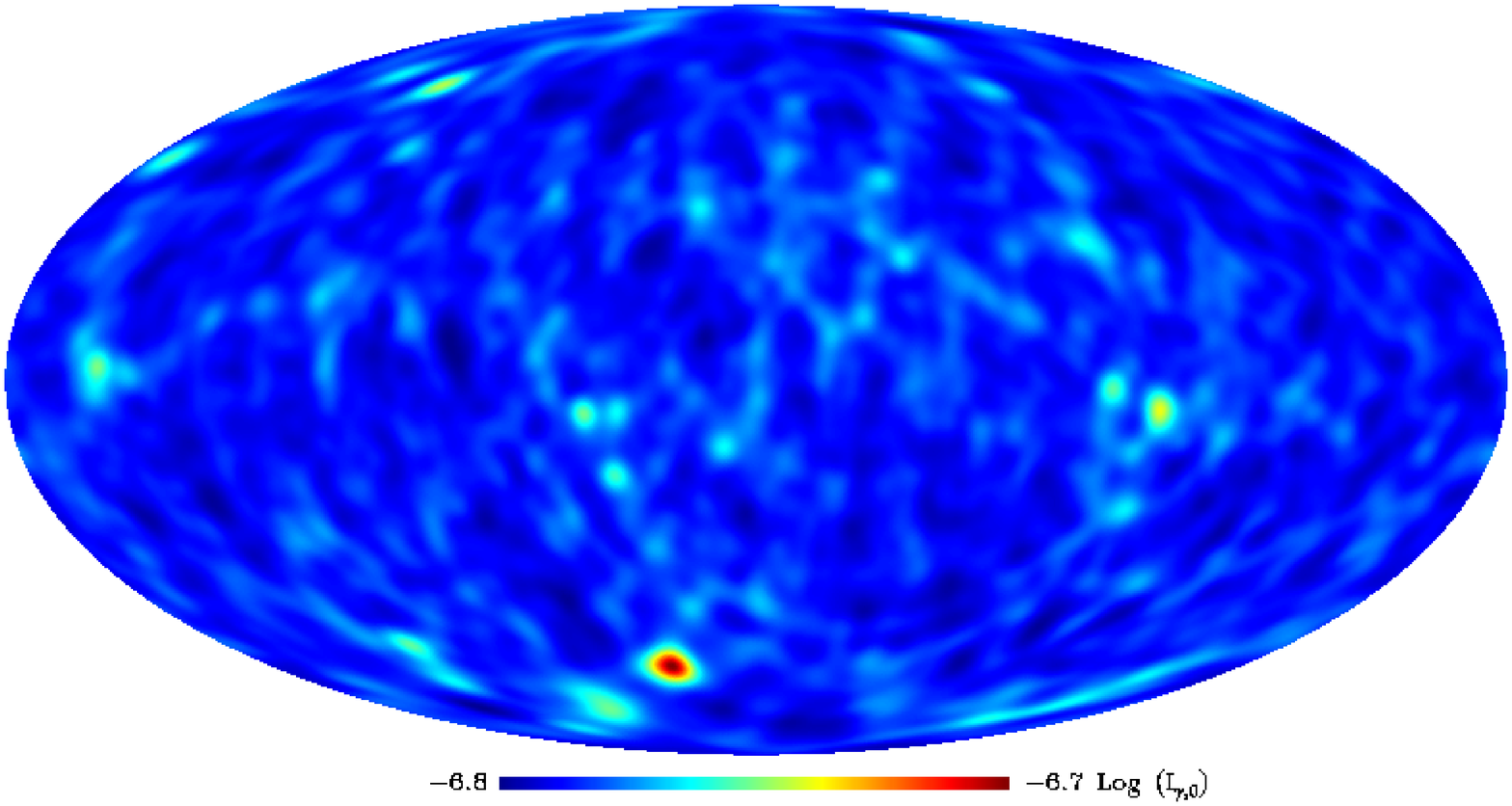}
\includegraphics[width=0.49\hsize]{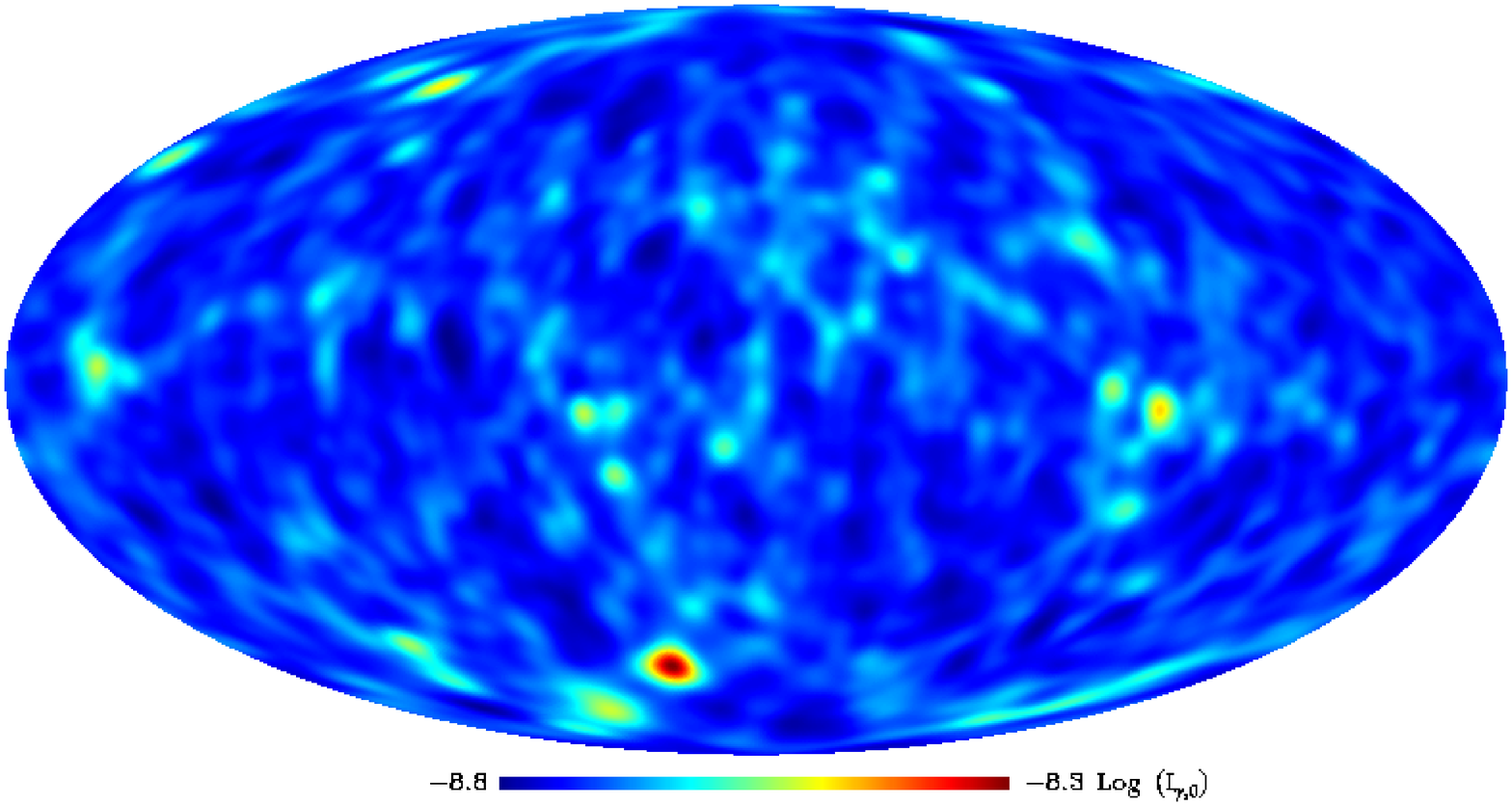}
\includegraphics[width=0.49\hsize]{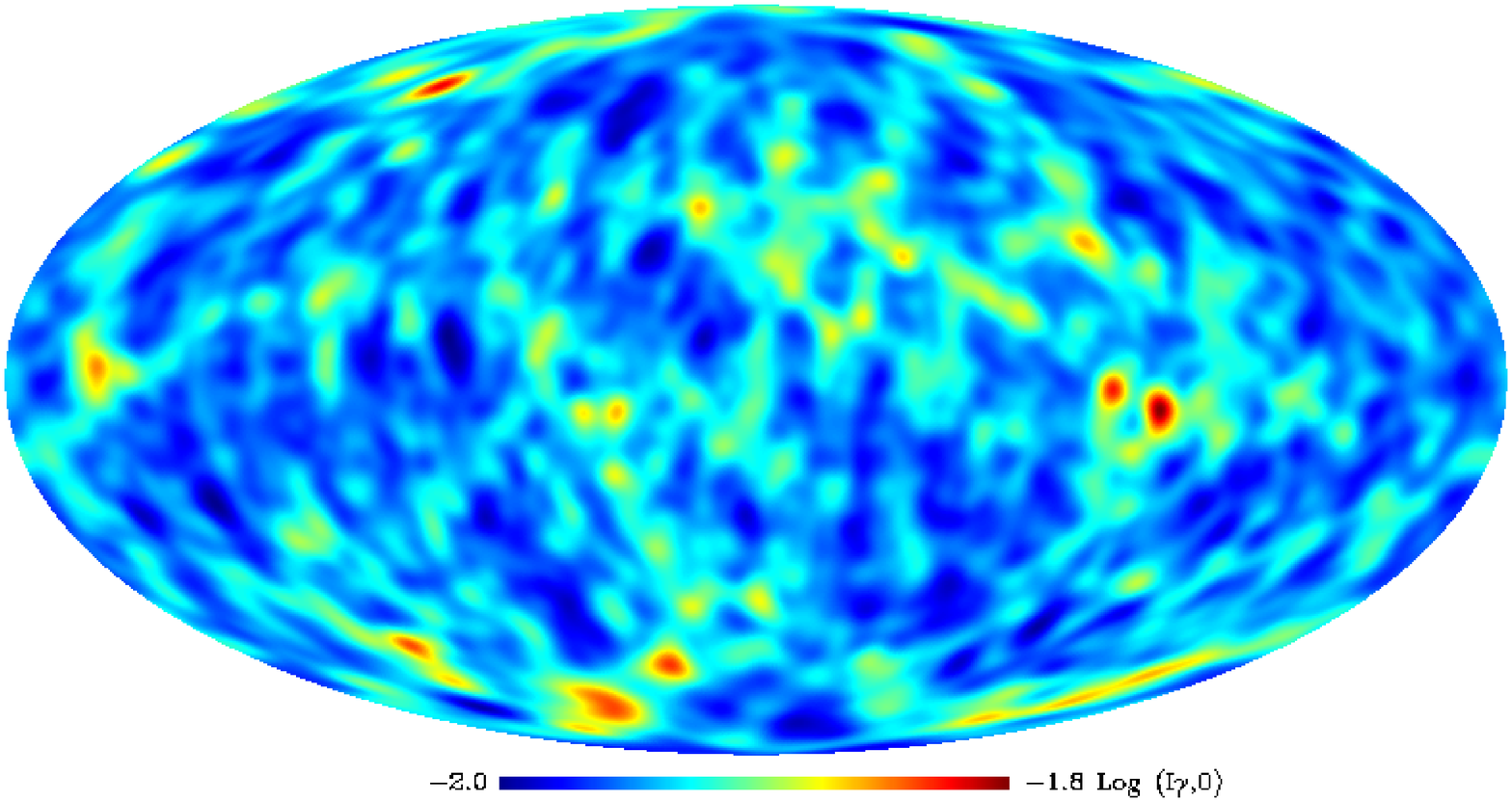}
\includegraphics[width=0.49\hsize]{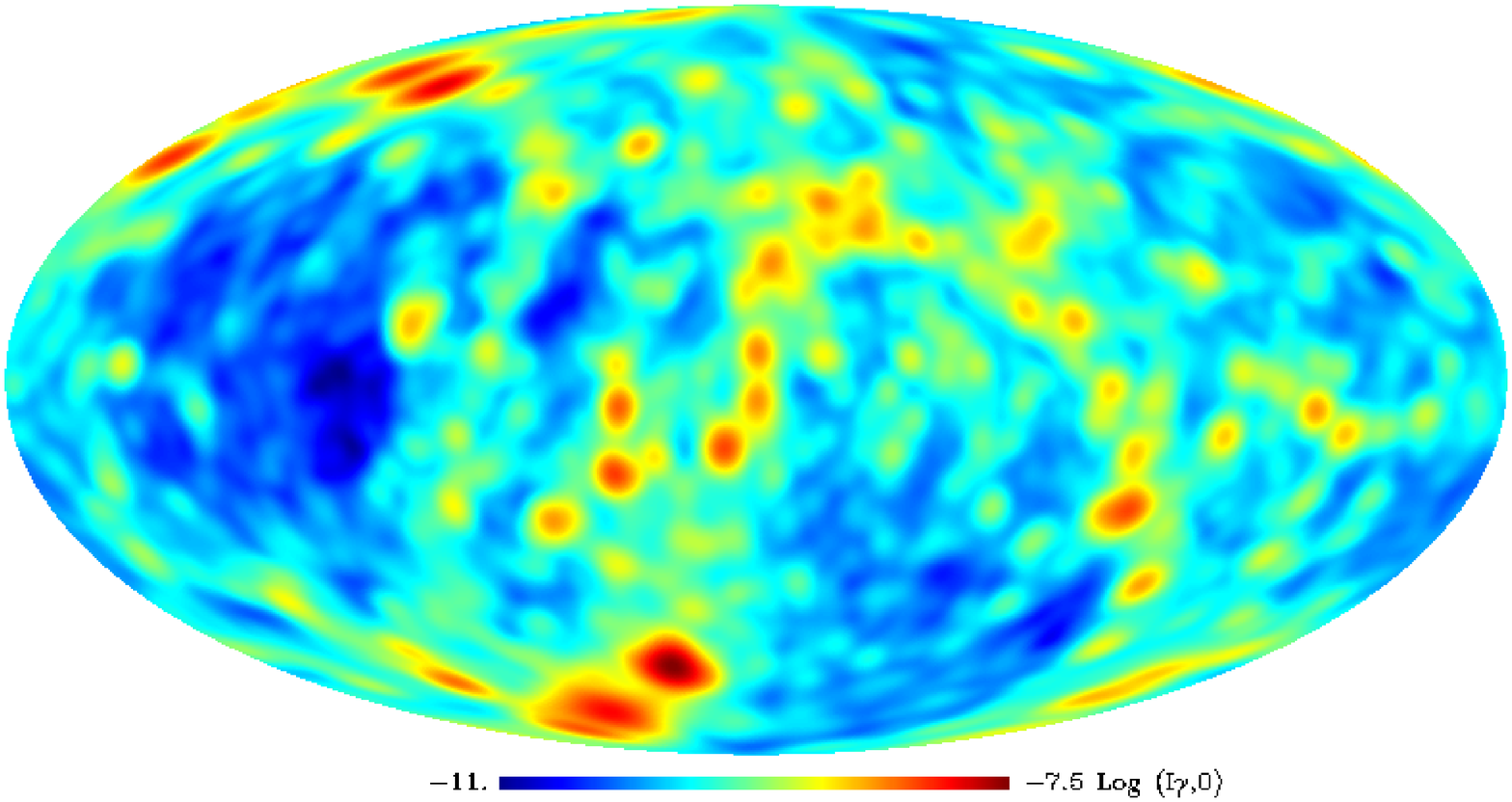}
\caption{{\footnotesize {\it Upper panel:} Full-sky maps at energies
  0.1 GeV and 32 GeV in
  the left and right, respectively.  The maps were smoothed with a
  Gaussian beam with a FWHM of $5^\circ$. At a single energy, a full-sky map
  is very smooth, nearby structures are only minimally visible. {\it Lower panel:} Ratio of
  the maps in the upper panel (left) and a partial map containing only nearby
  structures within 68 Mpc for an observed energy of 0.1 GeV
  (right). Creating difference maps (``color'' maps) using different energy channels greatly
  enhances the signal of nearby structures.}}
\label{smooth}
\end{figure*} 

\subsection{Energy dependence}

An important result that we obtain from the analysis of our results
is that there are differences in the shape of the power spectra
for different energies.  We can exploit these
differences and enhance the power at small or large angular scales by taking
the ratio of sky maps at different energies. The upper panel of Fig.~\ref{smooth}
shows the total sky maps for $E_{\gamma,0}=0.1\,{\rm GeV}$ and
$E_{\gamma,0}=32\,{\rm GeV}$ in the left and right, respectively, smoothed
with a Gaussian beam with a FWHM of $5^\circ$. Both maps show only small
anisotropies, with the exception of a couple of prominent structures. The sky
map in the lower left panel is the ratio of the two maps; it clearly enhances
the signal of nearby structures. Convincing supporting evidence for this
enhancement comes from the 
nearest partial map at $z=0$ for $E_{\gamma,0}=0.1\,{\rm GeV}$ shown in
the lower right corner of Fig.~\ref{smooth}.  Most of the prominent dark
matter structures that can be seen in this map are also clearly present in the
map with the energy ratio.  We conclude that the $\gamma$-ray sky maps that
the FERMI satellite will obtain at different energies could be used to construct difference
maps which may then show enhanced correlations with nearby cosmic large-scale
structure. 

\bibliography{lit}

\begin{thebibliography}{24}
\expandafter\ifx\csname natexlab\endcsname\relax\def\natexlab#1{#1}\fi
\expandafter\ifx\csname bibnamefont\endcsname\relax
  \def\bibnamefont#1{#1}\fi
\expandafter\ifx\csname bibfnamefont\endcsname\relax
  \def\bibfnamefont#1{#1}\fi
\expandafter\ifx\csname citenamefont\endcsname\relax
  \def\citenamefont#1{#1}\fi
\expandafter\ifx\csname url\endcsname\relax
  \def\url#1{\texttt{#1}}\fi
\expandafter\ifx\csname urlprefix\endcsname\relax\def\urlprefix{URL }\fi
\providecommand{\bibinfo}[2]{#2}
\providecommand{\eprint}[2][]{\url{#2}}

\bibitem[{\citenamefont{{Baltz} and {Edsj{\"o}}}(1999)}]{Baltz-Edsjo-99}
\bibinfo{author}{\bibfnamefont{E.~A.} \bibnamefont{{Baltz}}} \bibnamefont{and}
  \bibinfo{author}{\bibfnamefont{J.}~\bibnamefont{{Edsj{\"o}}}},
  \bibinfo{journal}{\prd} \textbf{\bibinfo{volume}{59}},
  \bibinfo{pages}{023511} (\bibinfo{year}{1999}).

\bibitem[{\citenamefont{{Berezinsky} et~al.}(1996)\citenamefont{{Berezinsky},
  {Bottino}, {Ellis}, {Fornengo}, {Mignola}, and {Scopel}}}]{Berezinsky-96}
\bibinfo{author}{\bibfnamefont{V.}~\bibnamefont{{Berezinsky}}}
  \bibinfo{author}{\bibnamefont{{et al.}}},
  \bibinfo{journal}{Astroparticle Physics} \textbf{\bibinfo{volume}{5}},
  \bibinfo{pages}{333} (\bibinfo{year}{1996}).

\bibitem[{\citenamefont{{Bertone} and {Merritt}}(2005)}]{Bertone-Merritt-05}
\bibinfo{author}{\bibfnamefont{G.}~\bibnamefont{{Bertone}}} \bibnamefont{and}
  \bibinfo{author}{\bibfnamefont{D.}~\bibnamefont{{Merritt}}},
  \bibinfo{journal}{Modern Physics Letters A} \textbf{\bibinfo{volume}{20}},
  \bibinfo{pages}{1021} (\bibinfo{year}{2005}).

\bibitem[{\citenamefont{{Jacholkowska}
  et~al.}(2006)\citenamefont{{Jacholkowska}, {Lamanna}, {Nuss}, {Bolmont},
  {Adloff}, {Alcaraz}, {Battiston}, {Brun}, and {et al.,}}}]{Jacholkowska-06}
\bibinfo{author}{\bibfnamefont{A.}~\bibnamefont{{Jacholkowska}}}
  \bibinfo{author}{\bibnamefont{{et al.}}}, \bibinfo{journal}{\prd}
  \textbf{\bibinfo{volume}{74}}, \bibinfo{pages}{023518}
  (\bibinfo{year}{2006}).

\bibitem[{\citenamefont{{Wood} et~al.}(2008)\citenamefont{{Wood}, {Blaylock},
  {Bradbury}, {Buckley}, {Byrum}, {Chow}, {Cui}, {de la Calle Perez}, and {et
  al.,}}}]{Wood-08}
\bibinfo{author}{\bibfnamefont{M.}~\bibnamefont{{Wood}}}
\bibinfo{author}{\bibnamefont{{et al.}}},
  \bibinfo{journal}{\apj} \textbf{\bibinfo{volume}{678}}, \bibinfo{pages}{594}
  (\bibinfo{year}{2008}).

\bibitem[{\citenamefont{{Stoehr} et~al.}(2003)\citenamefont{{Stoehr}, {White},
  {Springel}, {Tormen}, and {Yoshida}}}]{Stoehr-03}
\bibinfo{author}{\bibfnamefont{F.}~\bibnamefont{{Stoehr}}}
  \bibinfo{author}{\bibnamefont{{et al.}}},
  \bibinfo{journal}{\mnras} \textbf{\bibinfo{volume}{345}},
  \bibinfo{pages}{1313} (\bibinfo{year}{2003}).

\bibitem[{\citenamefont{{Springel} et~al.}(2008)\citenamefont{{Springel},
  {White}, {Frenk}, {Navarro}, {Jenkins}, {Vogelsberger}, {Wang}, {Ludlow}, and
  {et al.,}}}]{Springel-08}
\bibinfo{author}{\bibfnamefont{V.}~\bibnamefont{{Springel}}}
  \bibinfo{author}{\bibnamefont{{et al.}}}, \bibinfo{journal}{\nat}
  \textbf{\bibinfo{volume}{456}}, \bibinfo{pages}{73} (\bibinfo{year}{2008}).

\bibitem[{\citenamefont{{Strong} et~al.}(2004)\citenamefont{{Strong},
  {Moskalenko}, and {Reimer}}}]{Strong-Moskalenko-Reimer-04}
\bibinfo{author}{\bibfnamefont{A.~W.} \bibnamefont{{Strong}}},
  \bibinfo{author}{\bibfnamefont{I.~V.} \bibnamefont{{Moskalenko}}},
  \bibnamefont{and} \bibinfo{author}{\bibfnamefont{O.}~\bibnamefont{{Reimer}}},
  \bibinfo{journal}{\apj} \textbf{\bibinfo{volume}{613}}, \bibinfo{pages}{956}
  (\bibinfo{year}{2004}).

\bibitem[{\citenamefont{{Atwood} et~al.}(2009)\citenamefont{{Atwood}, {Abdo},
  {Ackermann}, {Althouse}, {Anderson}, {Axelsson}, {Baldini}, {Ballet}, and {et
  al.,}}}]{Atwood-09}
\bibinfo{author}{\bibfnamefont{W.~B.} \bibnamefont{{Atwood}}}
  \bibinfo{author}{\bibnamefont{{et al.}}}, \bibinfo{journal}{\apj}
  \textbf{\bibinfo{volume}{697}}, \bibinfo{pages}{1071} (\bibinfo{year}{2009}).

\bibitem[{\citenamefont{{Ando} et~al.}(2007{\natexlab{a}})\citenamefont{{Ando},
  {Komatsu}, {Narumoto}, and {Totani}}}]{Ando-07a}
\bibinfo{author}{\bibfnamefont{S.}~\bibnamefont{{Ando}}}
 \bibinfo{author}{\bibnamefont{{et al.}}},
  \bibinfo{journal}{\mnras} \textbf{\bibinfo{volume}{376}},
  \bibinfo{pages}{1635} (\bibinfo{year}{2007}{\natexlab{a}}).

\bibitem[{\citenamefont{{Jubelgas} et~al.}(2008)\citenamefont{{Jubelgas},
  {Springel}, {En{\ss}lin}, and {Pfrommer}}}]{Jubelgas-08}
\bibinfo{author}{\bibfnamefont{M.}~\bibnamefont{{Jubelgas}}}
  \bibinfo{author}{\bibnamefont{{et al.}}},
  \bibinfo{journal}{\aap} \textbf{\bibinfo{volume}{481}}, \bibinfo{pages}{33}
  (\bibinfo{year}{2008}).

\bibitem[{\citenamefont{{Ando} and {Komatsu}}(2006)}]{Ando-Komatsu-06}
\bibinfo{author}{\bibfnamefont{S.}~\bibnamefont{{Ando}}} \bibnamefont{and}
  \bibinfo{author}{\bibfnamefont{E.}~\bibnamefont{{Komatsu}}},
  \bibinfo{journal}{\prd} \textbf{\bibinfo{volume}{73}},
  \bibinfo{pages}{023521} (\bibinfo{year}{2006}).

\bibitem[{\citenamefont{{Siegal-Gaskins} and
  {Pavlidou}}(2009)}]{Siegal-Gaskins-Pavlidou-09}
\bibinfo{author}{\bibfnamefont{J.~M.} \bibnamefont{{Siegal-Gaskins}}}
  \bibnamefont{and}
  \bibinfo{author}{\bibfnamefont{V.}~\bibnamefont{{Pavlidou}}},
  \bibinfo{journal}{Physical Review Letters} \textbf{\bibinfo{volume}{102}},
  \bibinfo{pages}{241301} (\bibinfo{year}{2009}).

\bibitem[{\citenamefont{{Zavala} et~al.}(2009)\citenamefont{{Zavala},
  {Springel}, and {Boylan-Kolchin}}}]{Zavala-Springel-Boylan-Kolchin-09}
\bibinfo{author}{\bibfnamefont{J.}~\bibnamefont{{Zavala}}},
  \bibinfo{author}{\bibfnamefont{V.}~\bibnamefont{{Springel}}},
  \bibnamefont{and}
  \bibinfo{author}{\bibfnamefont{M.}~\bibnamefont{{Boylan-Kolchin}}},
  \bibinfo{journal}{ArXiv e-prints}  (\bibinfo{year}{2009}),
  \eprint{arXiv:0908.2428}.

\bibitem[{\citenamefont{{Boylan-Kolchin}
  et~al.}(2009)\citenamefont{{Boylan-Kolchin}, {Springel}, {White}, {Jenkins},
  and {Lemson}}}]{Boylan-Kolchin-09}
\bibinfo{author}{\bibfnamefont{M.}~\bibnamefont{{Boylan-Kolchin}}}
\bibinfo{author}{\bibnamefont{{et al.}}},
  \bibinfo{journal}{\mnras} \textbf{\bibinfo{volume}{398}},
  \bibinfo{pages}{1150}  (\bibinfo{year}{2009}).

\bibitem[{\citenamefont{{Feng}}(2005)}]{Feng-05}
\bibinfo{author}{\bibfnamefont{J.~L.} \bibnamefont{{Feng}}},
  \bibinfo{journal}{Annals of Physics} \textbf{\bibinfo{volume}{315}},
  \bibinfo{pages}{2} (\bibinfo{year}{2005}).

\bibitem[{\citenamefont{{Battaglia} et~al.}(2004)\citenamefont{{Battaglia},
  {Roeck}, {Ellis}, {Gianotti}, {Olive}, and {Pape}}}]{Battaglia-04}
\bibinfo{author}{\bibfnamefont{M.}~\bibnamefont{{Battaglia}}}
\bibinfo{author}{\bibnamefont{{et al.}}},
  \bibinfo{journal}{European Physical Journal C} \textbf{\bibinfo{volume}{33}},
  \bibinfo{pages}{273} (\bibinfo{year}{2004}).

\bibitem[{\citenamefont{{Gondolo} et~al.}(2004)\citenamefont{{Gondolo},
  {Edsj{\"o}}, {Ullio}, {Bergstr{\"o}m}, {Schelke}, and {Baltz}}}]{Gondolo-04}
\bibinfo{author}{\bibfnamefont{P.}~\bibnamefont{{Gondolo}}}
 \bibinfo{author}{\bibnamefont{{et al.}}}, \bibinfo{journal}{JCAP} \textbf{\bibinfo{volume}{7}}, \bibinfo{pages}{8}
  (\bibinfo{year}{2004}).

\bibitem[{\citenamefont{{Gondolo} et~al.}(2005)\citenamefont{{Gondolo},
  {Edsj{\"o}}, {Ullio}, {Bergstrom}, {Schelke}, and {Baltz}}}]{Gondolo-05}
\bibinfo{author}{\bibfnamefont{P.}~\bibnamefont{{Gondolo}}}
\bibinfo{author}{\bibnamefont{{et al.}}}, \bibinfo{journal}{New Astronomy Review}
  \textbf{\bibinfo{volume}{49}}, \bibinfo{pages}{149} (\bibinfo{year}{2005}).

\bibitem[{\citenamefont{{Bringmann} et~al.}(2008)\citenamefont{{Bringmann},
  {Bergstr{\"o}m}, and {Edsj{\"o}}}}]{Bringmann-Bergstrom-Edsjo-08}
\bibinfo{author}{\bibfnamefont{T.}~\bibnamefont{{Bringmann}}},
  \bibinfo{author}{\bibfnamefont{L.}~\bibnamefont{{Bergstr{\"o}m}}},
  \bibnamefont{and}
  \bibinfo{author}{\bibfnamefont{J.}~\bibnamefont{{Edsj{\"o}}}},
  \bibinfo{journal}{Journal of High Energy Physics}
  \textbf{\bibinfo{volume}{1}}, \bibinfo{pages}{49} (\bibinfo{year}{2008}).

\bibitem[{\citenamefont{{Heister} et~al.}(2004)\citenamefont{{Heister},
  {Schael}, {Barate}, {Bruneli{\`e}re}, {de Bonis}, {Decamp}, {Goy},
  {J{\'e}z{\'e}quel}, and {et al.,}}}]{Heister-04}
\bibinfo{author}{\bibfnamefont{A.}~\bibnamefont{{Heister}}}
\bibinfo{author}{\bibnamefont{{et al.}}},
  \bibinfo{journal}{Physics Letters B} \textbf{\bibinfo{volume}{583}},
  \bibinfo{pages}{247} (\bibinfo{year}{2004}).

\bibitem[{\citenamefont{{N{\'u}{\~n}ez}
  et~al.}(2008)\citenamefont{{N{\'u}{\~n}ez}, {Zavala}, {Nellen}, {Sussman},
  {Cabral-Rosetti}, {Mondrag{\'o}n}, {Instituto Avanzado de Cosmolog{\'{\i}}a},
  and {IAC}}}]{Nunez-08}
\bibinfo{author}{\bibfnamefont{D.}~\bibnamefont{{N{\'u}{\~n}ez}}}
  \bibinfo{author}{\bibnamefont{{et al.}}}, \bibinfo{journal}{JCAP} \textbf{\bibinfo{volume}{5}}, \bibinfo{pages}{3}
  (\bibinfo{year}{2008}).

\bibitem[{\citenamefont{{Navarro} et~al.}(1997)\citenamefont{{Navarro},
  {Frenk}, and {White}}}]{Navarro-Frenk-White-97}
\bibinfo{author}{\bibfnamefont{J.~F.} \bibnamefont{{Navarro}}},
  \bibinfo{author}{\bibfnamefont{C.~S.} \bibnamefont{{Frenk}}},
  \bibnamefont{and} \bibinfo{author}{\bibfnamefont{S.~D.~M.}
  \bibnamefont{{White}}}, \bibinfo{journal}{\apj}
  \textbf{\bibinfo{volume}{490}}, \bibinfo{pages}{493} (\bibinfo{year}{1997}).

\bibitem[{\citenamefont{{Ando} et~al.}(2007{\natexlab{b}})\citenamefont{{Ando},
  {Komatsu}, {Narumoto}, and {Totani}}}]{Ando-07b}
\bibinfo{author}{\bibfnamefont{S.}~\bibnamefont{{Ando}}}
\bibinfo{author}{\bibnamefont{{et al.}}},
  \bibinfo{journal}{\prd} \textbf{\bibinfo{volume}{75}},
  \bibinfo{pages}{063519} (\bibinfo{year}{2007}{\natexlab{b}}).

\end{thebibliography}

%\begin{thebibliography}{99} 

%\end{thebibliography}

\end{document}